\documentclass[aps,prl,preprint,groupedaddress]{revtex4-1}

\def\be{\begin{equation}}
\def\ee{\end{equation}}
\def\beq{\begin{eqnarray}}
\def\eeq{\end{eqnarray}}
\def\n{\nonumber}
\usepackage{graphicx}

\begin{document}


\title{Qualitative analysis of the  Tolman metrics within the unimodular framework}


\author{Sudan Hansraj}

\altaffiliation{}
\affiliation{Astrophysics and Cosmology Research Unit, University of KwaZulu Natal}
\email[]{hansrajs@ukzn.ac.za}
\author{Njabulo Mkhize}

\altaffiliation{} \affiliation{ UKZN} \email[]{mkhizen18@gmail.com}

\date{\today}

\begin{abstract}

We investigate the behaviour of the Tolman metrics within the
formalism of the trace-free (or unimodular) gravity.  While
this approach is similar to the standard Einstein field equations,
some subtlety arises. The effective number of independent  field
equations is reduced by one on account of the density and pressure
appearing as an inseparable entity the inertial mass density.
Further energy is not conserved within the trace-free theory but the
conservation law may be supplemented to the field equations. This presentation of the field equations
offers a different avenue to determine the density and pressure
explicitly. It turns out that an extra integration constant is always in
evidence. While this constant has little impact on the dynamics and energy
conditions, it makes a significant impact on the gravitational mass
and equation of state. Graphical plots are generated to analyse the behaviour of physical quantities qualitatively.

\end{abstract}

\pacs{}

\maketitle

\section{Introduction}

Unimodular gravity,  also known as trace-free Einstein gravity,
originated from Einstein himself as a method to simplify the
analysis of the field equations of general relativity by fixing a
coordinate system with a constant volume element. The idea was
revived by Weinberg \cite{weinberg} as a potential paradigm to
explain the phenomenon of vacuum energy. Then the proposal lay
dormant until Ellis \cite{ellis1,ellis2} realised that the large
discrepancy in the value of the cosmological constant predicted by
quantum field theory and that of observation and measurement may be
explained by invoking the trace-free field equations. In this
formulation the cosmological constant is reduced to merely a
constant of integration instead of an innocuous object inserted by
hand to address the accelerated expansion of the universe problem.
Moreover, Ellis \cite{ellis1} demonstrated that for compact objects
the usual Isreal--Darmois boundary conditions are preserved. Note
that unimodular gravity and the Einstein standard theory are
completely equivalent.  Further important treatments of unimodular gravity may be found in \cite{unimod1,unimod2,unimod3}. 

Because the equations of motion are a
coupled system of up to ten partial differential equations, the
method of finding exact solutions is important. It has been shown by
Hansraj {\it{et al}} \cite{hans} that unravelling the Einstein
equations through the trace-free paradigm, offered a different
solution generating algorithm. Indeed the old solutions are still
valid, however, the solution methods sometimes yield more general
behaviour in the solution. For example, the exterior metric in the
unimodular scenario should be the Schwarzschild metric by the
Jebsen--Birkhoff theorem. The theorem asserts that the Schwarzschild
solution is a consequence strictly of spherical symmetry and
independent of whether the distribution is static or not. Through
the trace-free algorithm a de-Sitter term in $r^2$, $r$ being the
radial parameter, appears in the solution of the differential
equation. The implication is that although the cosmological constant
has been hidden in the field equations, it still lives on in the
solution space.

In view of the foregoing, if indeed the unimodular framework
constitutes a viable theory of gravity,  it is important to ask what
the behaviour of astrophysical compact objects would be like in this
scenario. Usually the effect of the cosmological constant is
negligible if not zero when constructing models of stars. However,
in this formulation, the hidden cosmological constant has the
potential to alter the physics as was demonstrated by Hansraj {\it
{et al}} \cite{hans}. While the trace--free equations and the
standard Einstein field equations are equivalent, their presentation
as a system of nonlinear partial differential equations leave room
for variation in behaviour. Whereas in the standard theory, there
exists  a system of at most 10 differential equations in the space
and dynamical variables, there are  9 in the trace--free version
since the density and pressure are inextricably linked as the
inertial mass density $\rho + p$. In the Einstein equations, the law
of energy conservation arises from the vanishing divergence of the
energy momentum tensor and generates the equation of hydrodynamical
equilibrium. This equation conveys no new information compared to
the standard Einstein field equations and may substitute one of the
field equations. In contrast, in the trace--free theory the energy
conservation does not arise in the same way and must be inserted by
hand. That is the 9 field equations must be supplemented by the
conservation law for a complete system of equations governing the
gravitational field. This is where the potential for some variation
in the dynamics exists. In the work of Hansraj {\it{et al}}
\cite{hans} it was shown that the well known Finch--Skea \cite{fs} stellar
model was supplemented by additional terms which affected the
profiles of the energy density and pressure and consequently the
sound speed, stellar mass and surface redshift. The Finch--Skea model was shown to be consistent with the astrophysical theory of Walecka \cite{wal}.

Alternate or extended theories of gravity have aroused considerable
interest recently. The principal motivation to modify Einstein's
equations is the observed accelerated expansion  of the universe
that is not a consequence of the standard model. Therefore,
proposals to include higher derivative or higher curvature terms
have been invoked.  In particular, Einstein--Gauss--Bonnet (EGB)
theory has proved promising in this regard. Strong support for
involving the Gauss--Bonnet term lies in the fact that this term
appears in the effective low energy action of heterotic string
theory \cite{gross}.  The
Gauss--Bonnet term is the second order term in the more general
Lovelock polynomial \cite{lovelock1,lovelock2} which is constructed from terms polynomial in
the Ricci tensor, Ricci scalar and Riemann tensor. The Lovelock
action is the most general action generating at most second order
equations of motion. A drawback of this theory is that the dynamical
behaviour is only impacted for spacetime dimensions higher than 4
but the standard theory is regained for orders less than or equal to
4. In Starobinksy's \cite{staro} $f(R)$ theory  the action that is
proposed is a polynomial in the Ricci scalar. While this idea has
the potential to account for the late time accelerated expansion of
the universe, it suffers from the severe drawback of yielding
derivatives of orders higher than two (ghosts) in the equations of motion. It
is usually expected that gravitational behavior is characterized by
up to second order equations of motion and that the Newtonian theory
would be regained in the appropriate limit.  Recently $f(R)$ theory
has been shown to be conformal to scalar tensor theory. 

In this work we analyze the  effect of trace--free gravity on exact
 solutions found by Tolman \cite{tolman} from his study of static spherically symmetric perfect fluid field equations.
 The Tolman metrics were derived after writing the equation of
 pressure isotropy in the special form
 \begin{equation}
    \frac{d}{dr}\left(\frac{e^{-\lambda}-1}{r^2}\right)+\frac{d}{dr}\left(\frac{e^{-\lambda}\nu'}{2r}\right)+e^{-\lambda-\nu}\frac{d}{dr}\left(\frac{e^{\nu}\nu'}{2r}\right)=0.
\end{equation}
 As the system of partial
 differential equations is underdetermined, choices for one of the
 metric potentials were made on the basis of the vanishing of some
 of the terms in the isotropy equation in the form given by Tolman.
 In each of Tolman solutions, we review the original assumptions and the consequent metric potentials.
 In the dynamical quantities  found by Tolman, we then proceed to insert the Tolman metric components into our
  trace--free algorithm in order to probe the dynamical quantities.

\vspace{0,5cm}
\section{Trace--Free Field Equations}

In order to facilitate a direct comparison with the work of Tolman,
we follow his conventions. The static spherically symmetric
spacetime in coordinates $(t, r, \theta, \phi)$ is taken as
\begin{equation}
ds^2 = e^{\nu(r)}dt^2  - e^{\lambda (r)} dr^2 - r^2 \left(d\theta^2
+\sin^2 \theta d\phi^2\right) \label{1}
\end{equation}
where the gravitational potentials $\nu$ and $\lambda$ are functions
of the radial coordinate $r$ only. We utilise a comoving fluid
4-velocity $ u^a = e^{-\nu /2} \delta_0^a $ a perfect fluid source
with energy momentum tensor $T_{ab} = (\rho + p)u_a u_b - p g_{ab}$
in geometrized units setting the gravitational constant $G$ and the
speed of light $c$ to unity. The quantities $\rho$ and $p$ are the
energy density and pressure respectively.

The trace--free Einstein field equations are given by 
\begin{equation}
R_{ab} - \frac{1}{4} Rg_{ab}=T_{ab}-\frac{1}{4}Tg_{ab}  \label{2y}
\end{equation}
where $T$ is the trace of the energy momentum tensor.  The
trace--free components of the energy-momentum tensor  are given by
\begin{equation}
\hat{T}_{ab} = \left( \frac{3}{4}(\rho + p)e^{2\nu},
\frac{1}{4}(\rho + p)e^{2\lambda}, \frac{r^2}{4} (\rho + p),
\frac{r^2 \sin^2 \theta}{4} (\rho + p)\right) \label{3}
\end{equation}
from which the coupling of density ($\rho$) and pressure ($p$) is
readily apparent. We follow the notation of \cite{ellis1} and the hat symbol refers to trace-less quantities. Ordinarily in the regular Einstein field equations
the $G_{tt}$ and $T_{tt}$ components are free of the pressure
variable.

The trace--free field equations may now be expressed as 
\begin{eqnarray}
(2\nu'' + \nu'^2   - \nu'\lambda' ) +\frac{4}{r}\left(\nu' +
\lambda'\right) +\frac{4}{r^2}(e^{\lambda}-1)
&=&6(\rho + p)e^{\lambda} \label{24a}\\ \n \\
\frac{4}{r}\left(\nu' + \lambda'\right)-(2\nu'' + \nu'^2   -
\nu'\lambda' ) -\frac{4}{r^2}(e^{\lambda}-1)
&=& 2(\rho + p)e^{\lambda}  \label{24b}\\ \n \\
(2\nu'' + \nu'^2   - \nu'\lambda' ) + \frac{4}{r^2}(e^{\lambda}-1)
&=& 2(\rho + p)e^{\lambda}. \label{24c}
\end{eqnarray}

These three equations are not independent. Subtracting three times
equation (\ref{24b}) from (\ref{24a}) and equating (\ref{24b}) and
(\ref{24c}) give the master set of field equations
\begin{eqnarray}\label{2f11}
2\nu''+ \nu'^{2}-\nu'\lambda'- \frac{2}{r}(\nu'+ \lambda')+ \frac{4}{r^2}(e^{\lambda}-1)&=&0  \label{211a} \\ \n \\
\frac{\nu'+ \lambda'}{r}&= &(\rho + p)e^{\lambda} \label{211b}\\ \n \\
p' + \frac{1}{2}(\rho + p)\nu' &= &0 \label{211cc}
\end{eqnarray}
where the last equation (\ref{211cc}) is the conservation equation
$T^{ab}_{\,\,;b} =0$. Note that the conservation law is necessary in
the trace--free field equations since the divergence of $\hat{T}_{ab}$
 does not vanish in general.

While the trace--free version of the field equations are equivalent
to the standard Einstein system, the presentation as a system of
partial differential equations is manifestly different. This
presentation raises the question whether more general behaviour in
the known metrics may be found on solving the system of equations.
The equation of pressure isotropy (\ref{24a}) is identical to the
standard version so any metric known to solve the standard Einstein
equations may be utilised. The trace--free system offers a useful
algorithm to find exact solutions.  Once a metric is selected, the
components may be substituted into (\ref{24b}) to find the inertial
mass density $\rho + p$. This quantity may then be substituted into
(\ref{24c}) to reveal the pressure profile explicitly. Finally
removing the pressure from the inertial mass density generates the
energy density. We shall implement this algorithm in what follows.

When constructing models of stellar distributions composed of
perfect fluid matter, the following conditions are usually imposed
in order that the model is physically reasonable. The energy density
($\rho$) and pressure ($p$) profiles are expected to be positive
definite with the pressure vanishing for some radial value that
demarcates the boundary of the fluid according to the
Israel--Darmois junction conditions.  Generally it is preferred that
both functions are monotonically decreasing from the centre
outwardly although this requirement may be too strict in compact
matter. The sound speed should be subluminal and obey the causality
criterion $0 < \frac{dp}{d\rho} < 1$. The interior metric must
smoothly match the exterior Schwarzschild solution across the
boundary hypersurface.  The energy conditions must be satisfied.
That is the (i)  weak energy condition: $\rho - p > 0$, (ii) strong
energy condition: $\rho + p > 0$ and (iii) dominant energy
condition: $\rho + 3p > 0$.  For static fluid spheres with a
monotonically decreasing and positive pressure profile, the surface
redshift $z_{r = R} =\frac{1}{\sqrt{-g_{00}}}-1 =e^{-\nu/2}-1$  should be
less than 2. The Buchdahl \cite{buch} limit  $ \frac{\rm
mass}{\rm radius} < \frac{4}{9}$ governing the mass-radius ratio
ensures the stability of the sphere must be satisfied.

\vspace{0,5cm}
\section{Tolman I metric (Einstein Universe)}

Following Einstein, Tolman began with  the assumption
$e^{\nu}=\text{const.}$ as the simplest prescription of a variable
to solve the field equations. The metric potentials are then found
to be $e^{\lambda}=\frac{1}{1-\frac{r^2}{R^2}}$ and $ e^{\nu}= c^2$
for some constants $c$ and $R$. Accordingly the dynamical quantities
work out to $\rho=\frac{3}{R^2}$ and $ p=-\frac{1}{R^2}$. In the trace--free situation the density and pressure are calculated
as $ \rho=\frac{2}{R^2}-K$ and $ p=K$. The energy  conditions $\rho
-p=\frac{2}{R^2} -2K$, $\rho +p=\frac{2}{R^2} $ and $\rho
+3p=\frac{2}{R^2} +2K$. This is the well--known static Einstein universe solution providing
a non-zero  energy density and pressure. The severe defect in this
model is the constant value of the density and pressure which does
not conform to observation. For this case, the density and pressure
 coincide with Tolman I so no new insight is gained.

\section{Tolman II metric (Schwarzschild--de Sitter)}

Based on his arrangement of the master field equation, Tolman made
the choice $e^{-\lambda-\nu}=\text{constant}$ and obtained the
potentials $
e^{\lambda}=\left(1-\frac{2m}{r}-\frac{r^2}{R^2}\right)^{-1} $ and $
e^{\nu}= c^2 \left(1-\frac{2m}{r}-\frac{r^2}{R^2}\right).$ The
density and pressure emerge as $ \rho=\frac{3}{R^2}$ and $
p=-\frac{3}{R^2}.$  Within the unimodular framework, the dynamics are $ \rho=-K$ and $
p=K$ for some constant $K$.   The above metrics lead to uniform
energy density and pressure as in the standard theory. The density
and pressure are constant and are related by the equation of state
$\rho +p=0$. This is a characteristic of dark matter -- a
speculative idea proposed to explain the observed accelerated
expansion of the universe. The expansion requires a negative
pressure and this is the case if $p=-\rho$. The term
$\left(1-\frac{2m}{r}\right)$, belongs to the Schwarzchild exterior
metric while the term quadratic in $r$ arises on account of the
cosmological constant which is believed to be the generator of
vacuum energy in empty space through curvature.

\section{Tolman III metric: Schwarzschild Interior}

Introducing the relationship $    e^{-\lambda}=1-\frac{r^2}{R^2}$,  the Tolman metric potentials evaluate to
\be    e^{\lambda} = \frac{1}{1-\frac{r^2}{R^2}} \hspace{0.5cm}{\mbox{and}}\hspace{0.5cm} e^\nu = \left[A-B\left(1-\frac{r^2}{R^2}\right)^{\frac{1}{2}}\right]^2 \n \ee
while the dynamical quantities have the form
$\rho =  \frac{3}{R^2}$ and $p = \frac{1}{R^2}\left[\frac{3B\sqrt{1-\frac{r^2}{R^2}}-A}{A-B\sqrt{1-\frac{r^2}{R^2}}}\right]$
where $A$, $B$ and $R$ are constants. The TFE algorithm generates
the same results. When $A=0$ the Tolman I metric is obtained. Note that the Schwarzschild potentials produce a constant density fluid. However, the converse is not necessarily true. In \cite{hans} it was demonstrated that beginning with the requirement of a constant density yielded metric potentials that included a Nariai term \cite{nariai}. When this term is suppressed the usual Schwarzschild metric is regained.

\vspace{0,5cm}
\section{Tolman IV metric}

This is the first of the new solutions of Tolman that exhibited previously unknown physical behaviour.
The assumption $\frac{e^{\nu}\nu'}{2r}=\text{const.} $ generated the
 metric potentials
\[
e^{\lambda}=\frac{1+\frac{2
r^2}{A^2}}{\left(1+\frac{r^2}{A^2}\right)
\left(1-\frac{r^2}{R^2}\right)} \hspace{0.5cm} {\mbox{and}}
\hspace{0.5cm}    e^{\nu}=B^2 \left(1+\frac{r^2}{A^2}\right)
\]
where $A$, $B$ and $R$ are constants. In the trace--free algorithm
the dynamical quantities are expressed as
\begin{eqnarray}\label{4es11}
\rho&=&\frac{\left(3 A^2+2 r^2\right) \left(A^2+2 R^2\right)}{2 R^2
\left(A^2+2 r^2\right)^2}-K\\ \nonumber \\ \label{4es12}
p&=&\frac{A^2+2 R^2}{2 R^2 \left(A^2+2 r^2\right)}+K.
\end{eqnarray}
These are equivalent to the Tolman quantities except for the
appearance of an extra constant of integration which indeed plays a
role in the dynamical evolution of the fluid. This is an artefact of
the trace--free equations.  The sound speed index has the form
\begin{eqnarray}
\frac{dp}{d\rho} =\frac{A^2+2 r^2}{5 A^2+2 r^2}  = \frac{1+2
\left(\frac{r}{A}\right)^2}{5+2 \left(\frac{r}{A}\right)^2}.
\end{eqnarray}
It is clear that $\frac{dp}{d\rho}\ge 0$ while demanding
$\frac{dp}{d\rho}\le 1$, leads to $\frac{4}{5+2
\left(\frac{r}{A}\right)^2}>0$, which is always time. Hence the
sound speed is subluminal for all values of
radii as well as constants within the Tolman IV metric.\\
The energy conditions work out to
\begin{eqnarray}
\rho -p&=&\frac{A^4 \left(1-2 K R^2\right)+2 A^2 R^2 \left(1-4 K r^2\right)-8 K r^4 R^2}{R^2 \left(A^2+2 r^2\right)^2}\\ \nonumber \\
\rho +p&=&\frac{2 \left(A^2+r^2\right) \left(A^2+2 R^2\right)}{R^2
\left(A^2+2 r^2\right)^2}\\ \nonumber \
\rho +3p&=&\frac{A^4 \left(2 K R^2+3\right)+A^2 \left(r^2 \left(8 K R^2+4\right)+6 R^2\right)+8 r^2 R^2 \left(K r^2+1\right)}{R^2 \left(A^2+2 r^2\right)^2}.\ \newline \\
\end{eqnarray}
and these may be studied with the aid of graphical plots. A
barotropic equation of state exists. Solving for $r^2$ in equation
(\ref{4es11}) and substituting in (\ref{4es12})  generates a
functional dependence of $p$ on $\rho$. This is a usual expectation
of perfect fluids and is given by
\begin{equation}
p(\rho)=\frac{2 \left(A^2+2 R^2\right) (K+\rho )}{\sqrt{\left(A^2+2 R^2\right) \left(A^2 \left(16 K R^2+16 \rho  R^2+1\right)+2 R^2\right)}+A^2+2 R^2}+K.\\ \newline \ \\
\end{equation}
The gravitational mass of the star is computed as
\begin{equation}
m(r)=\frac{r^3 \left(A^2 \left(3-2 K R^2\right)+2 R^2 \left(3-2 K
r^2\right)\right)}{6 R^2 \left(A^2+2 r^2\right)}
\end{equation}
in geometric units.\\
The compactification parameter expresses the ratio of mass to radius
throughout the distribution and has the form
\begin{equation}
    \Lambda=\frac{m(r)}{r}=\frac{r^2 \left(A^2 \left(3-2 K R^2\right)+2 R^2 \left(3-2 K r^2\right)\right)}{6 R^2 \left(A^2+2 r^2\right)}.
\end{equation}
while the gravitational surface redshift $ z =
\frac{1}{B\left[\sqrt{\left(\frac{r^2}{A^2}+1\right)}\right]}-1$
is expected to be less than 2 at the boundary $r = R$.

\begin{figure}
    \centering
 \includegraphics[width=6cm]{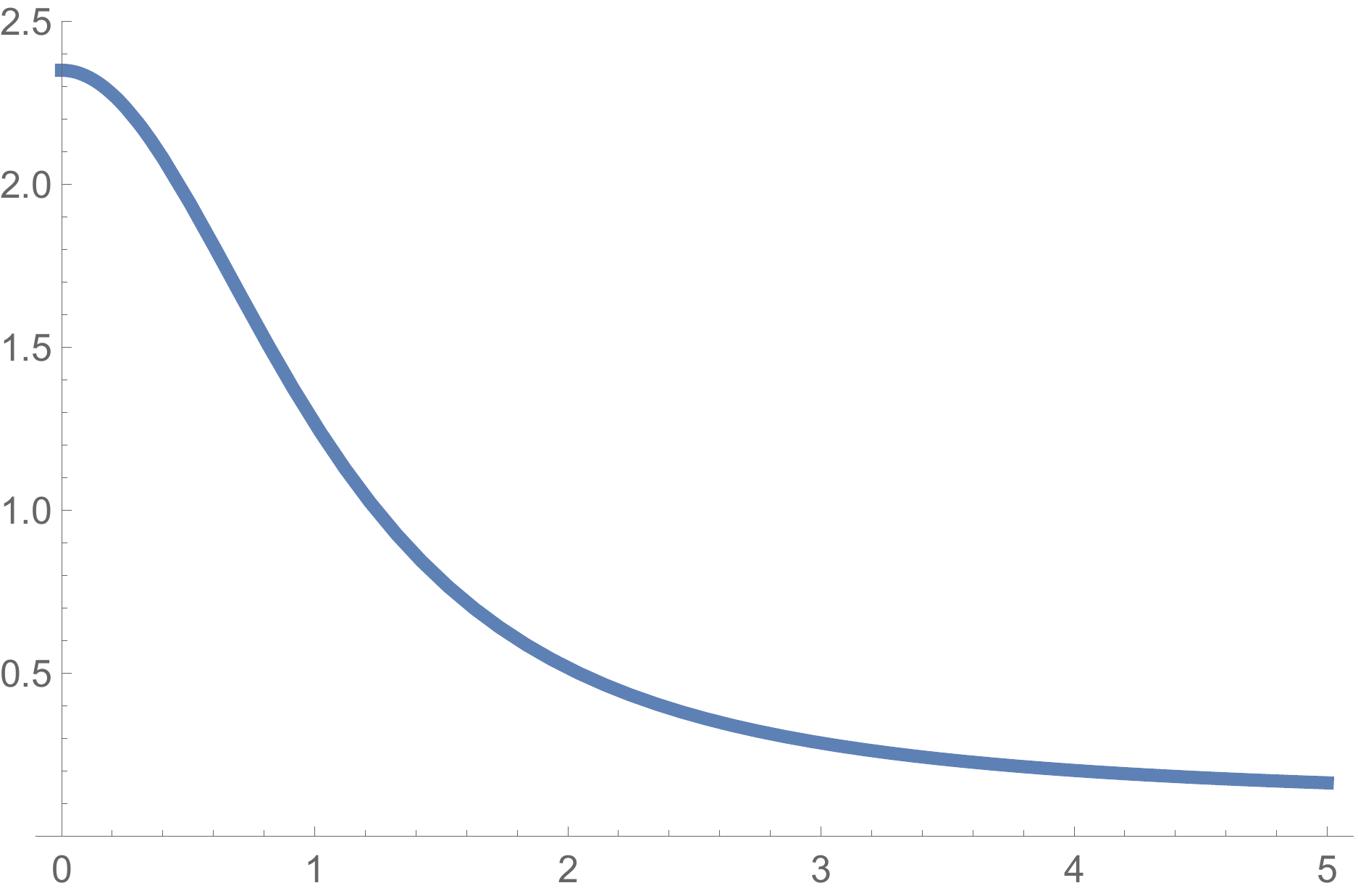}
    \caption{Plot of energy density ($\rho$) versus radius ($r$).}
\end{figure}

\newpage
\begin{figure}
    \centering
 \includegraphics[width=6cm]{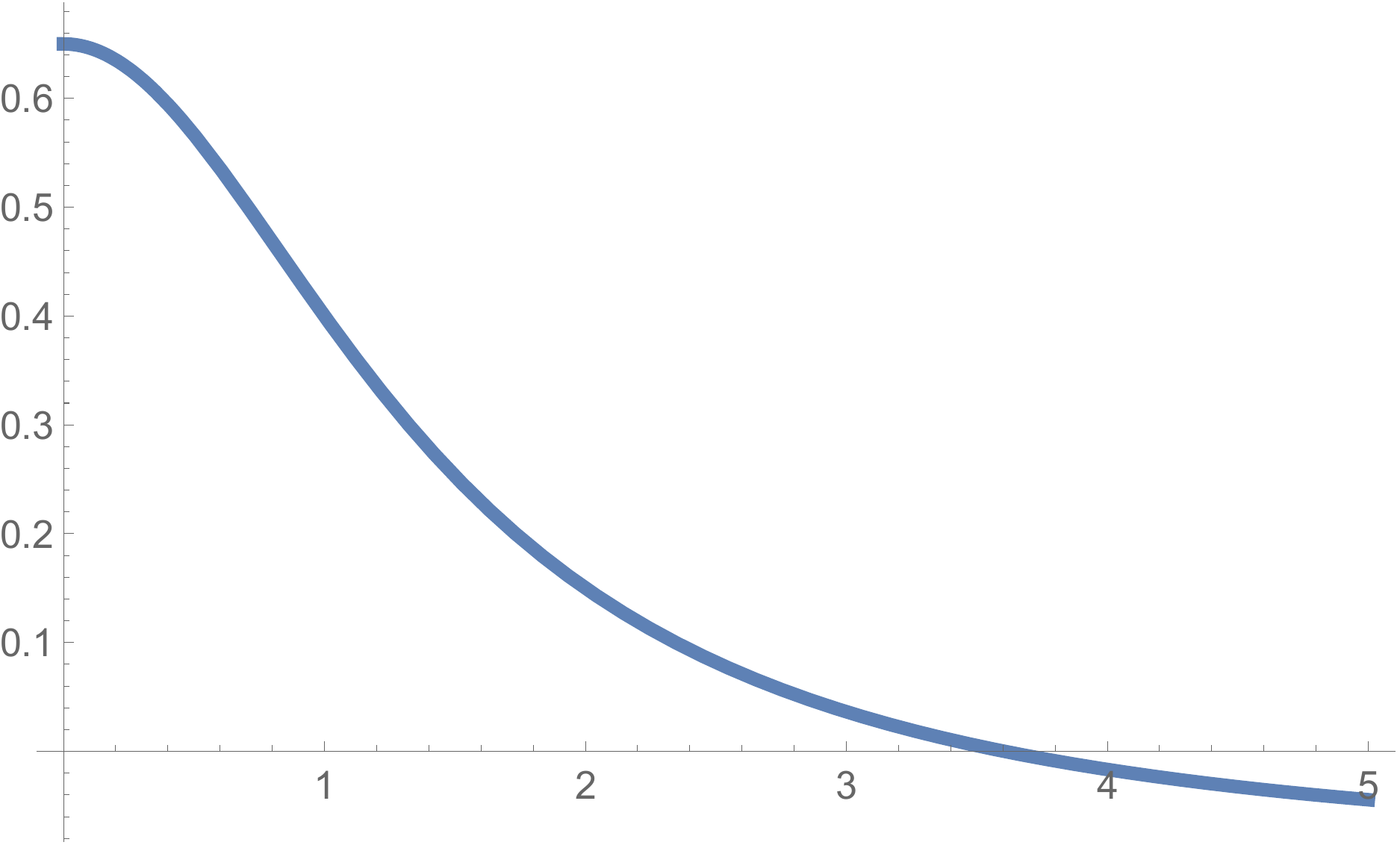}
    \caption{Plot of pressure ($p$) versus radius ($r$).}
\end{figure}

\newpage
\begin{figure}
    \centering
 \includegraphics[width=6cm]{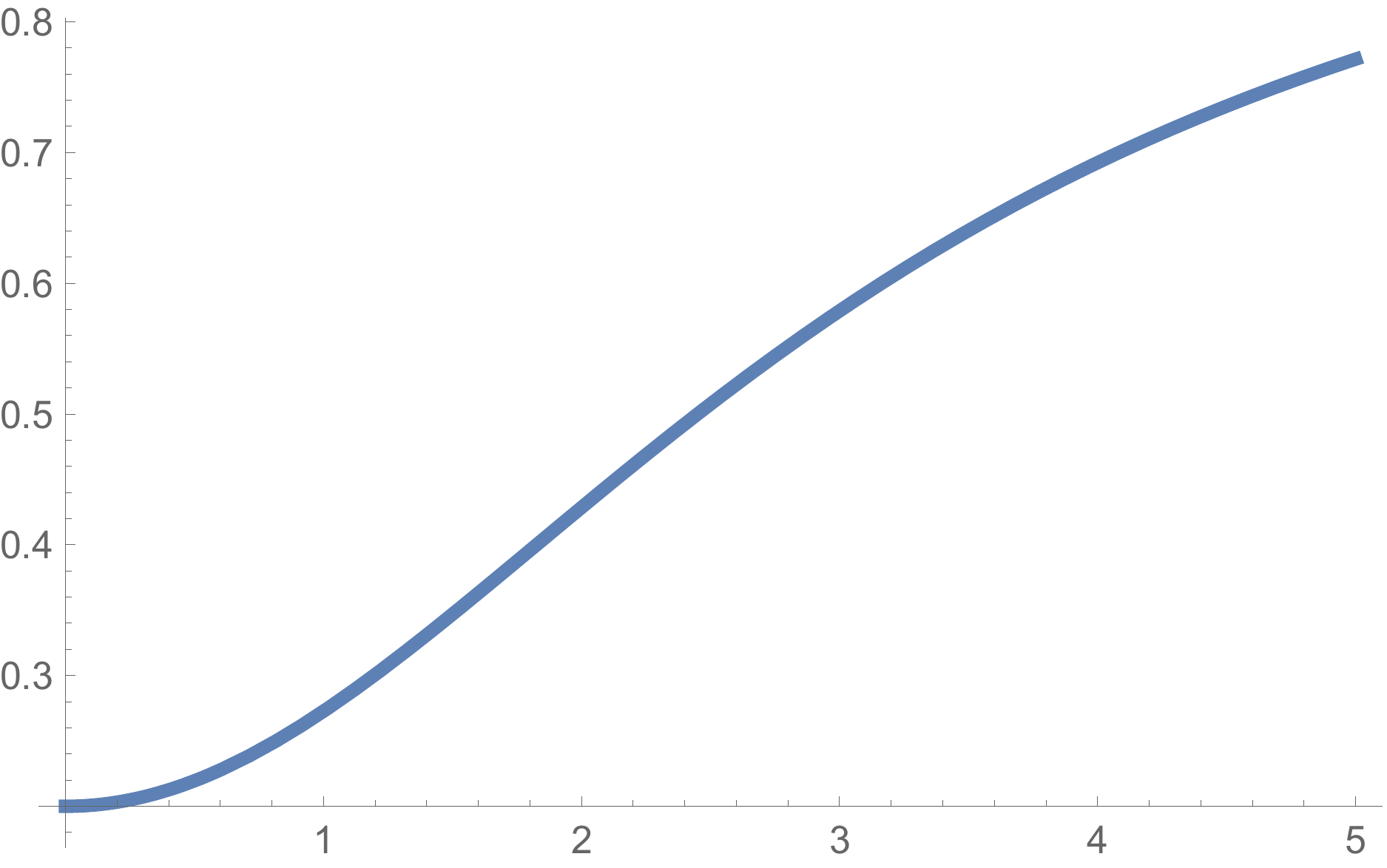}
    \caption{Plot of sound speed $\left(\frac{dp}{d\rho}\right)$ versus radius ($r$).}
\end{figure}

\newpage
\begin{figure}
    \centering
  \includegraphics[width=6cm]{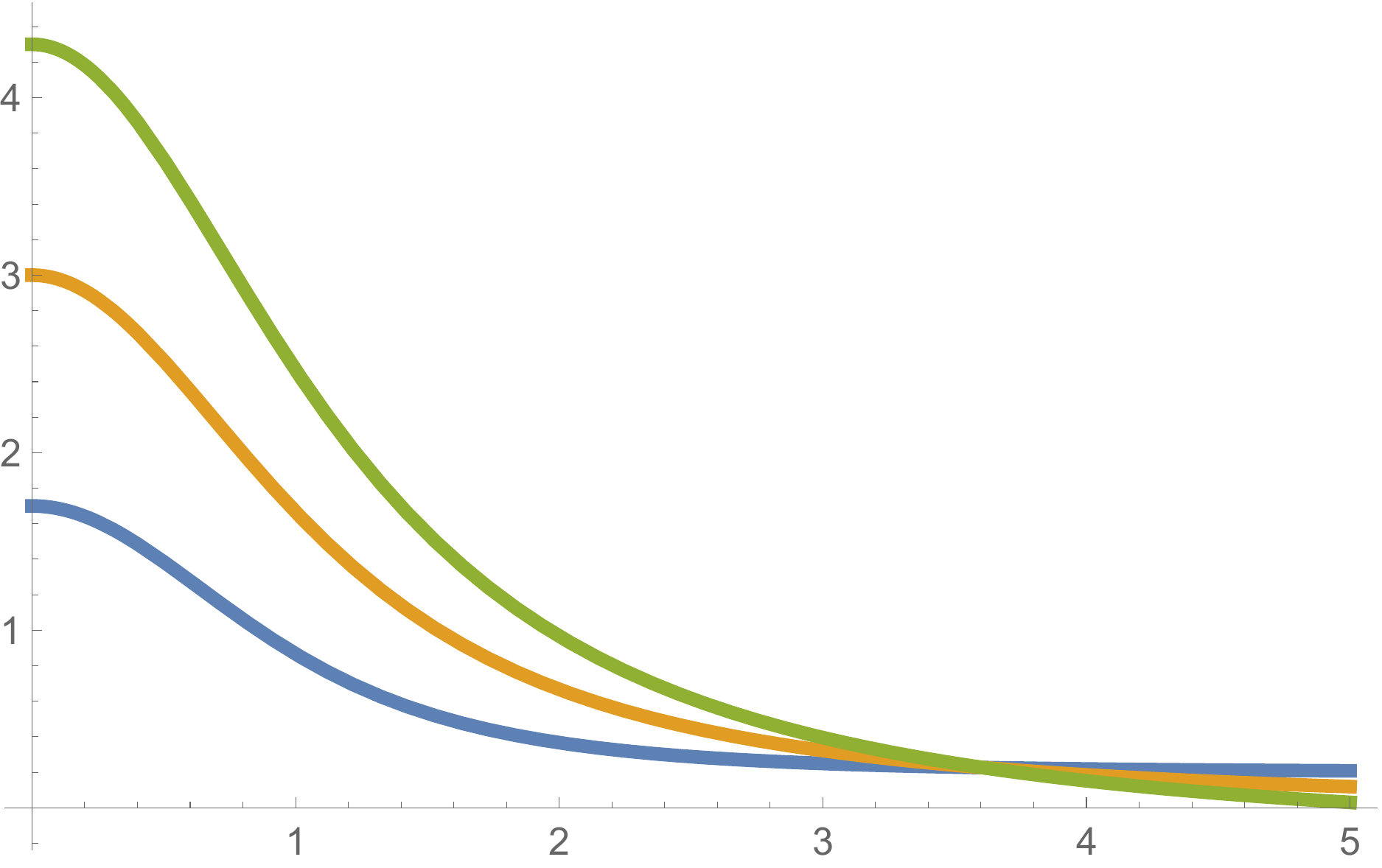}
    \caption{Plot of energy conditions ($\rho-p$, $\rho+p$ and $\rho+3p$) versus radius ($r$).}
\end{figure}

\newpage
\begin{figure}
    \centering
 \includegraphics[width=6cm]{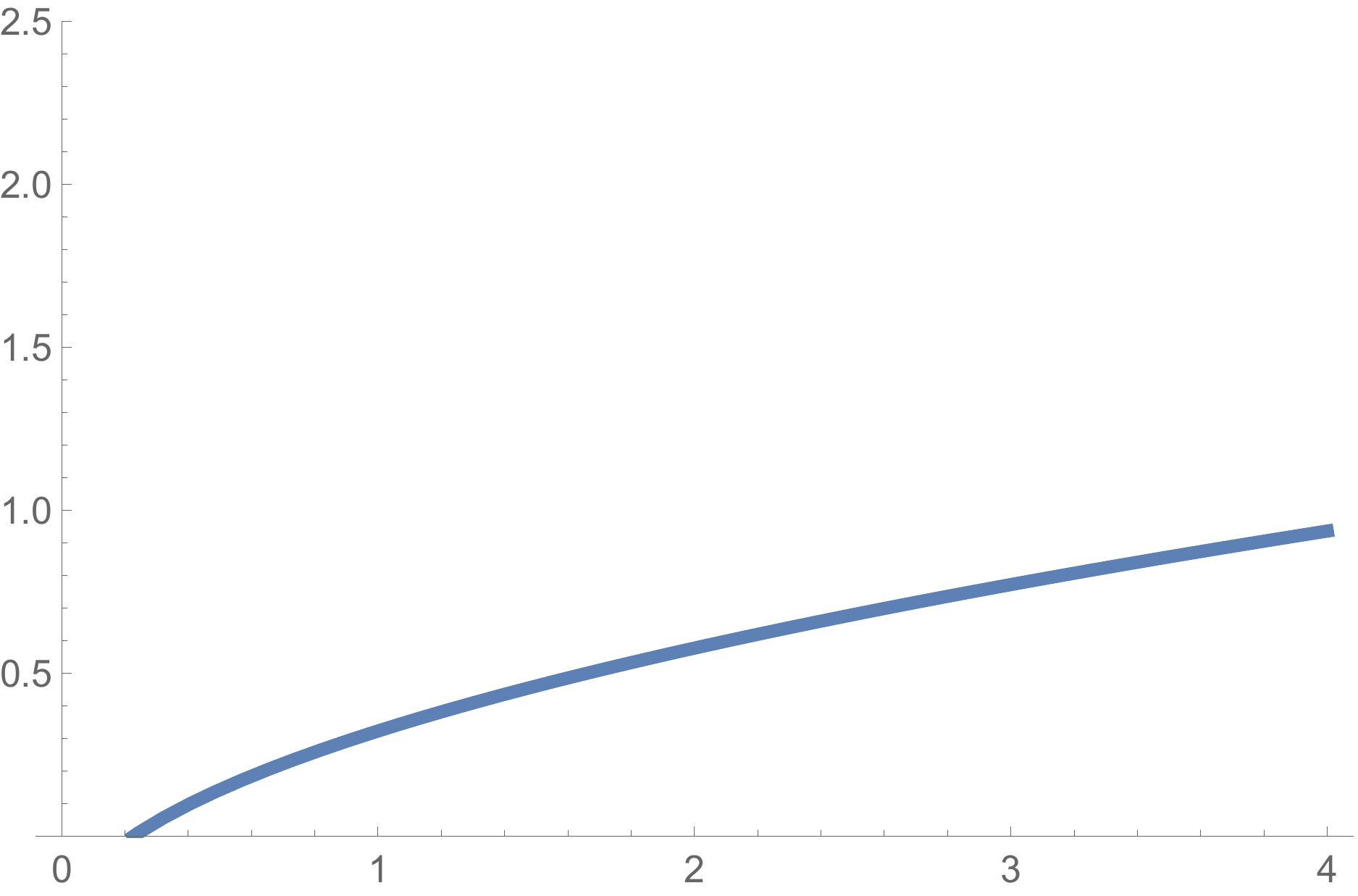}
    \caption{Plot of equation of state ($p=p(\rho)$) versus radius ($r$).}
\end{figure}

\newpage
\begin{figure}
    \centering
  \includegraphics[width=6cm]{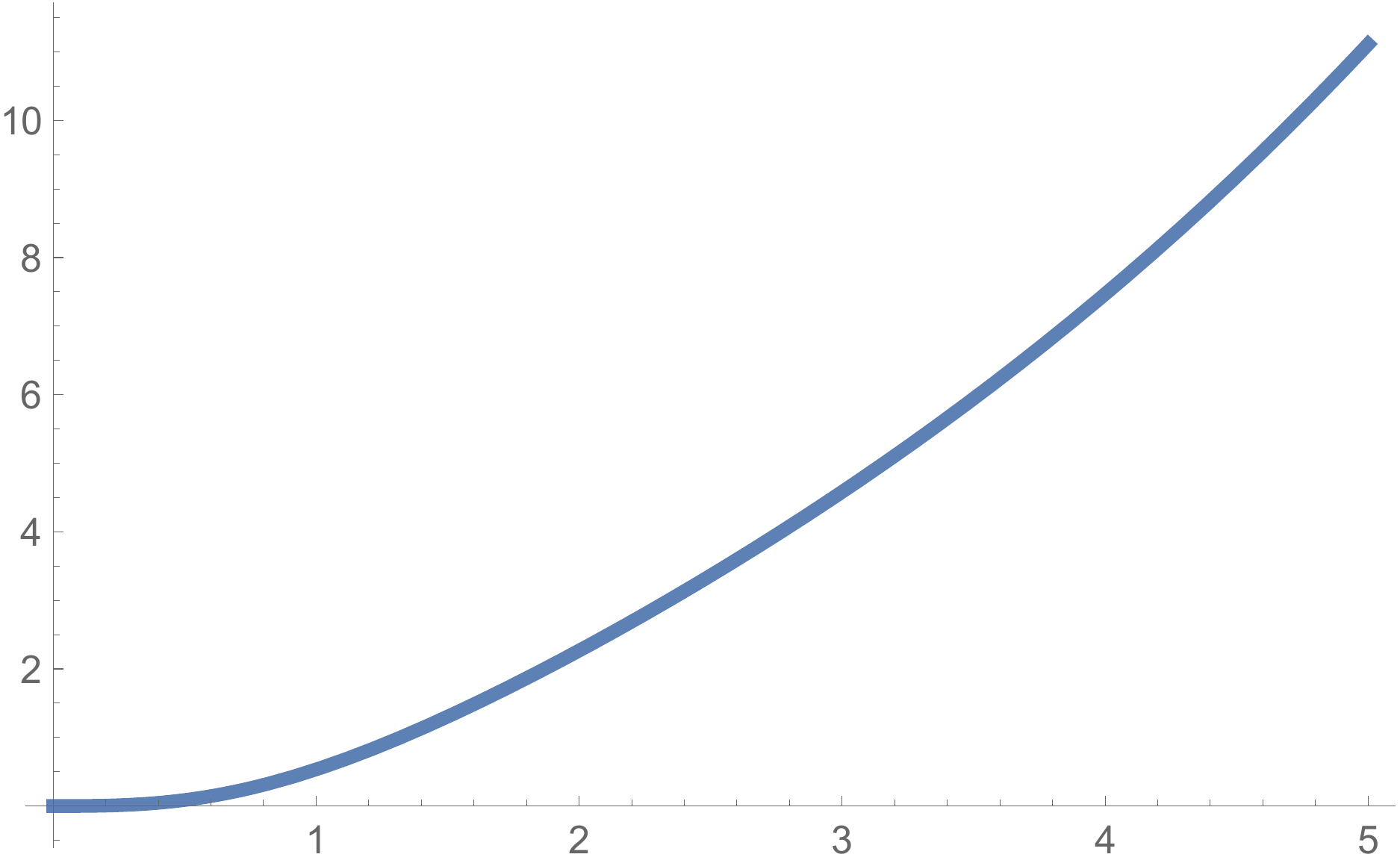}
    \caption{Plot of gravitational mass ($m(r)$) versus radius ($r$).}
\end{figure}

\newpage
\begin{figure}
    \centering
 \includegraphics[width=6cm]{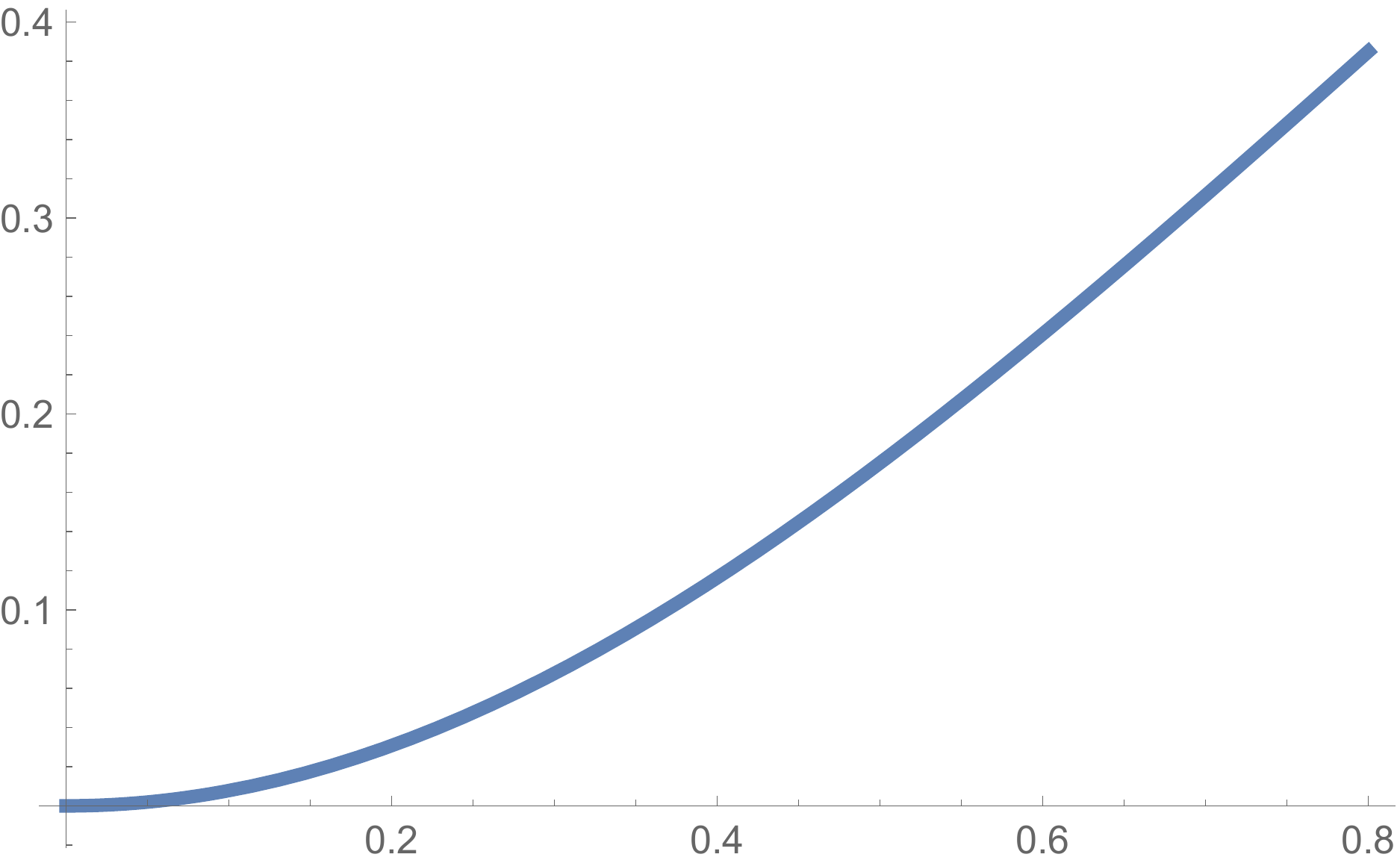}
    \caption{Plot of compactification parameter $(\Lambda)$ versus radius ($r$).}
\end{figure}

\newpage
{\bf{ Analysis of the plots:}} A very slight deviation on pressure
profile for this case is noted. Now for the purpose of analyzing our
model graphically we make the following parameter choices: $A=2$;
$B=0.5$; $K=-0.1$ and $R=1$.  Figure 1 portrays the behavior of the energy density
versus the radial coordinate ($r$). The plot clearly shows a
positive definite and a monotonically decreasing energy density with
increasing radius $r$ everywhere within the spherical distribution.
Figure 2 is the plot of the isotropic pressure versus radial $r$
coordinate. The plot evidently shows that the pressure is positive
at the origin, inside the boundary. At around radius $r=3.6$, it
vanishes indicating that the distribution has a finite radius.
Evidently, Figure 2 also demonstrates monotonic decrease with
increasing radius $r$ Figure 3 depicts  
the sound of speed value versus the radial  $r$ coordinate. The
profile satisfies the condition $0\le \frac{dp}{d\rho} \le1$, as
demanded for causality. Figure 4 exhibits the energy conditions
which are all positive inside the sphere. In figure 5 we provide a
plot for the equation of state, expressing the pressure as a
function of the density. It is a smooth singularity free function
within the star's radius. A plot of the gravitational mass is
displayed in figure 6. This reveals a smooth increasing function
with the increasing radius as is expected. The compactification
parameter (Figure 7) plot is an increasing function most
importantly satisfying the inequality
$\frac{\mbox{mass}}{\mbox{radius}}<\frac{4}{9}$. Finally, we
consider the  redshift profile (Figure 8) which is less than 2
units close to the radius $r=3.6$.  Therefore this model does satisfy the elementary
requirements for realistic behavior.

\section{Tolman V metric}

Tolman's prescription $e^{v} =\text{const.}r^{2n}$ resulted in the
potentials
\begin{eqnarray}
e^\lambda=\frac{1+2n-n^2}{1-\left(1+2n-n^2\right)
\left(\frac{r}{R}\right)^{N}} \hspace{0.5cm} {\mbox{and}}
\hspace{0.5cm}  e^{\nu}=B^2 r^{2n}
\end{eqnarray}
where $n$, $N=\frac{2 \left(1+2n-n^2\right)}{n+1}$ and $B$ are
constants. Setting $M={-\frac{2 \left(n^2+1\right)}{n+1}}$ the
trace--free algorithm yields
\begin{eqnarray}
\rho&=& \frac{(2-n)n}{(1+2n-n^2)r^2} - \frac{(2n+1)(n-3) r^2 \left(\frac{r}{R}\right)^M}{(n+1) R^4}-K\\ \nonumber \\
p&=&K-\frac{ (2 n+1) r^2 \left(\frac{r}{R}\right)^M}{R^4}+\frac{
n^2}{(1+2n-n^2) r^2}.
\end{eqnarray}
for the energy density and pressure respectively. Again we note that
the expressions are equivalent to Tolman however a significant
constant $K$ appears on account of the process of unravelling the
trace--free field equations. This must be a manifestation of the
cosmological constant that the trace--free equations sought to
conceal.  The sound speed is given by
\begin{equation}
\frac{dp}{d\rho} =\frac{(n-1) (n+1) (2 n+1) (1+2n-n^2) r^4-n (n+1)^2
R^4 \left(\frac{r}{R}\right)^M}{(n-3) (n-1) (2 n+1) (1+2n-n^2)
r^4-(n-2) (n+1)^2 R^4 \left(\frac{r}{R}\right)^M}.
\end{equation}
and the energy conditions are expressed as
\begin{eqnarray}
\rho -p&=&\frac{4 (2 n+1) r^2 \left(\frac{r}{R}\right)^M}{(n+1) R^4}-\frac{2 n ( n-1)}{(1+2n-n^2) r^2}-2 K\\ \nonumber \\
\rho +p&=&\frac{2 \left(1+n-2 n^2\right) r^2 \left(\frac{r}{R}\right)^M}{(n+1) R^4}+\frac{2 n}{(1+2n-n^2) r^2}\\ \nonumber \\
\rho +3p&=&2 K-\frac{4n (2n+1) r^2 \left(\frac{r}{R}\right)^M}{(n+1)
R^4}+\frac{2 n ( n+1)}{(1+2n-n^2) r^2}.
\end{eqnarray}
The gravitational mass function assumes the form
\begin{equation}
m(r)=\frac{r^5 \left(\frac{r}{R}\right)^{M}}{R^4}+\frac{(n-2) n
r}{n^2-2 n-1}-\frac{K r^3}{3}.
\end{equation}
while the compactification parameter is given by
\begin{equation}
\Lambda=\frac{r^4 \left(\frac{r}{R}\right)^{M}}{R^4}+\frac{(n-2) n
}{n^2-2 n-1}-\frac{K r^2}{3}
\end{equation}
The  redshift expression $ z=\frac{1}{Br^n}-1 $ must be
smaller than 2 at the boundary. In the main, the additive
integration constant $K$ acts to shift the values of the density,
pressure, sound speed and energy profiles. However, in the case of
the active mass a new term in $r^3$ is introduced and this
influences the overall mass of the star. Note that when $n=0$, the
density $\rho$ becomes constant and the result is reduced to the
Tolman I solution (Einstein universe) which was covered in the
previous chapter. In this solution Tolman decided to study the case
$n=\frac{1}{2}$ and to examine the physical properties of the
solution. Note that an equation of state may be explicitly obtained
depending on what value of $n$ is chosen.  To provide a wider
treatment of the behaviour of the model for various $n$ values we
select $n = -\frac{1}{2}$ (small dashes), $ n = 0$ (dotted line), $n
= \frac{1}{2}$ (dotted and dashed), $n = 1$ (thick), $n = 2$ (big
dashes)and $n = 3$ (solid thin curve). These values are suggested by
the expressions for the density and pressure since in each case some
$r$ dependence is suppressed.

\newpage
\begin{figure}
    \centering
\includegraphics[width=6cm]{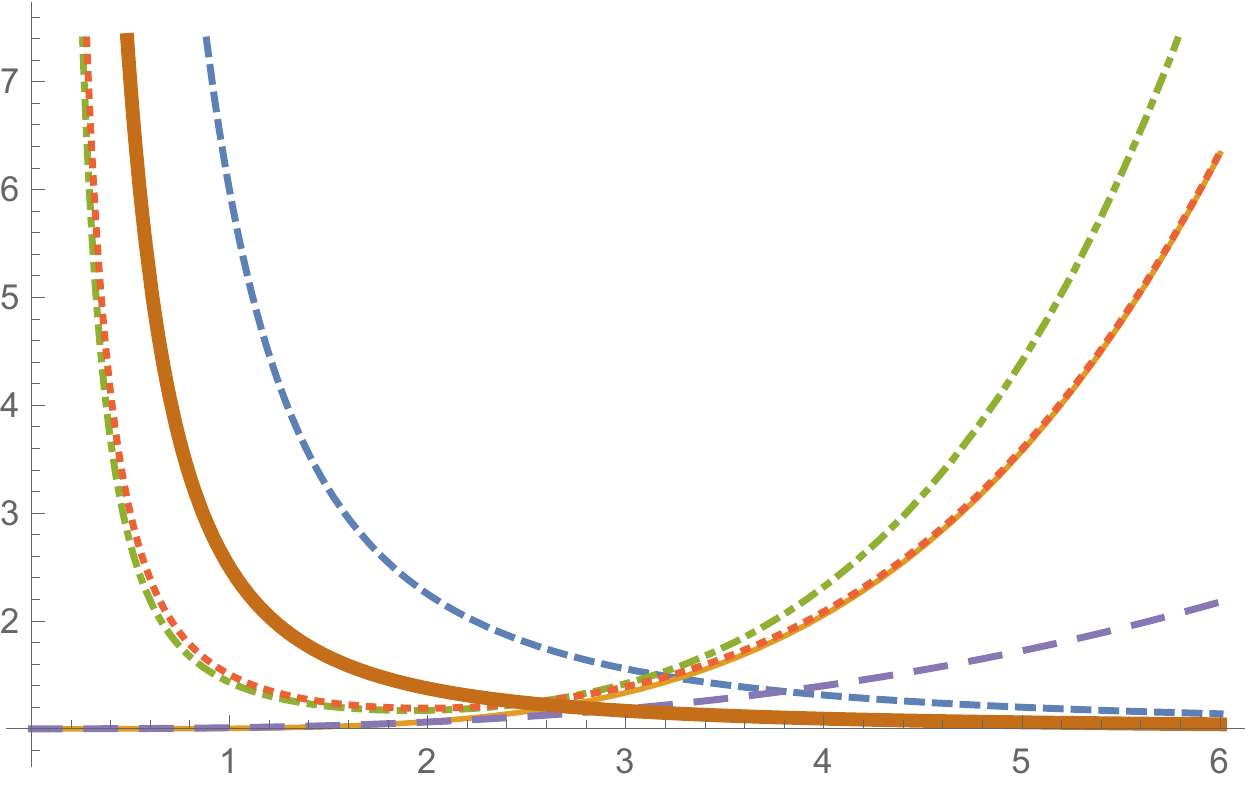}
    \caption{Energy density ($\rho$) versus radius ($r$).}
\end{figure}

\newpage
\begin{figure}
    \centering
  \includegraphics[width=6cm]{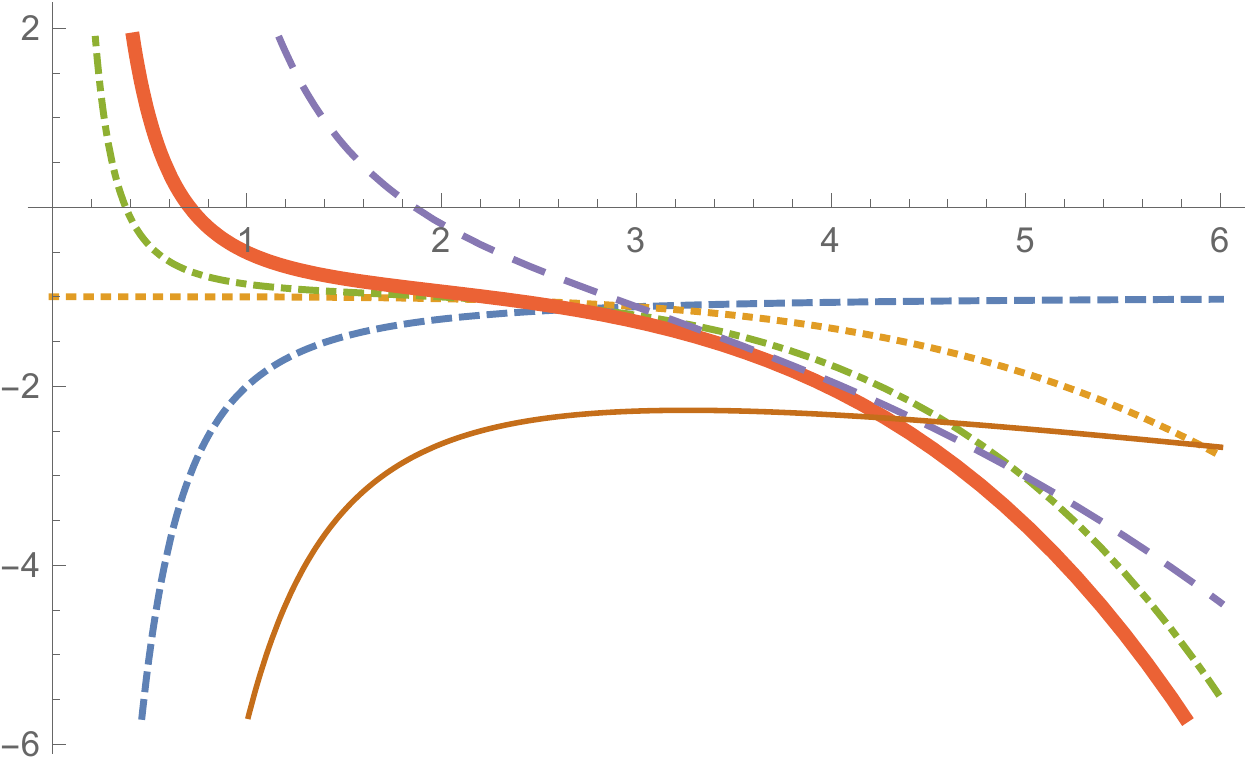}
    \caption{Pressure ($p$) versus radius ($r$).}
\end{figure}

\newpage
\begin{figure}
    \centering
 \includegraphics[width=6cm]{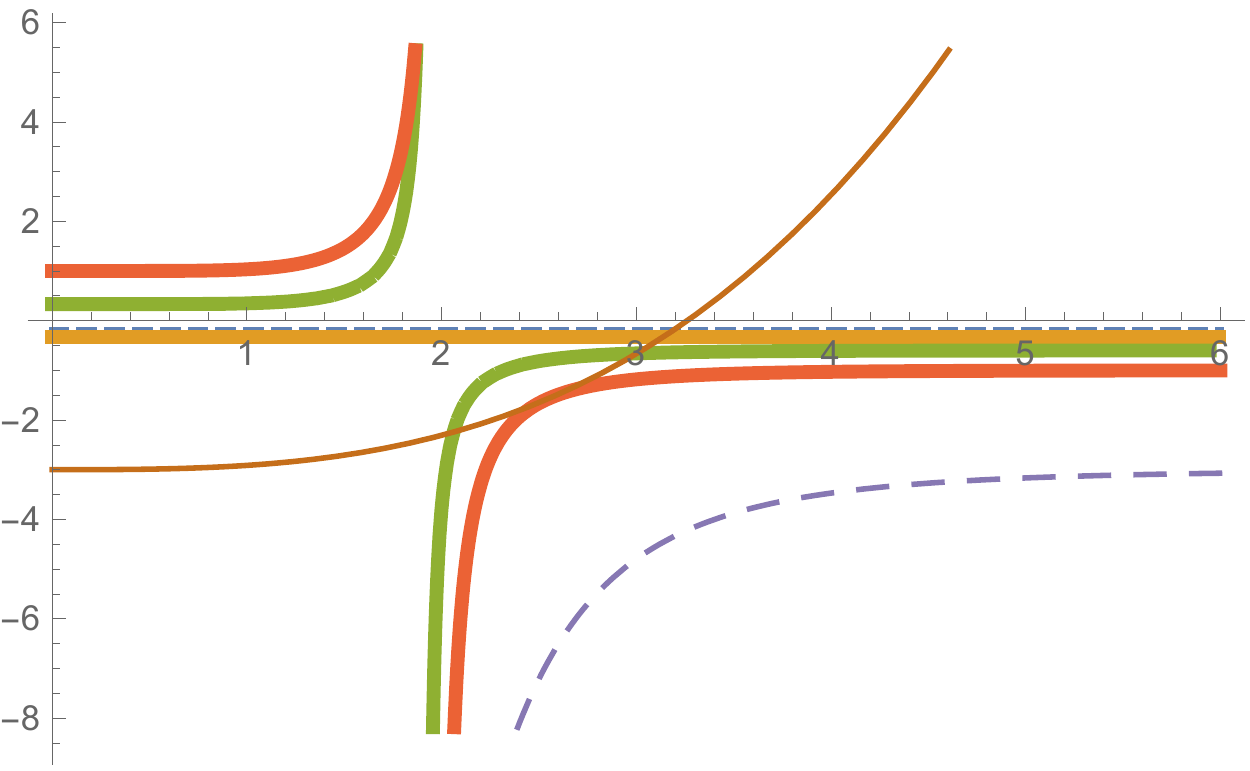}
    \caption{Sound speed index $\left(\frac{dp}{d\rho}\right)$ versus radius ($r$).}
\end{figure}

\newpage
\begin{figure}
    \centering
 \includegraphics[width=6cm]{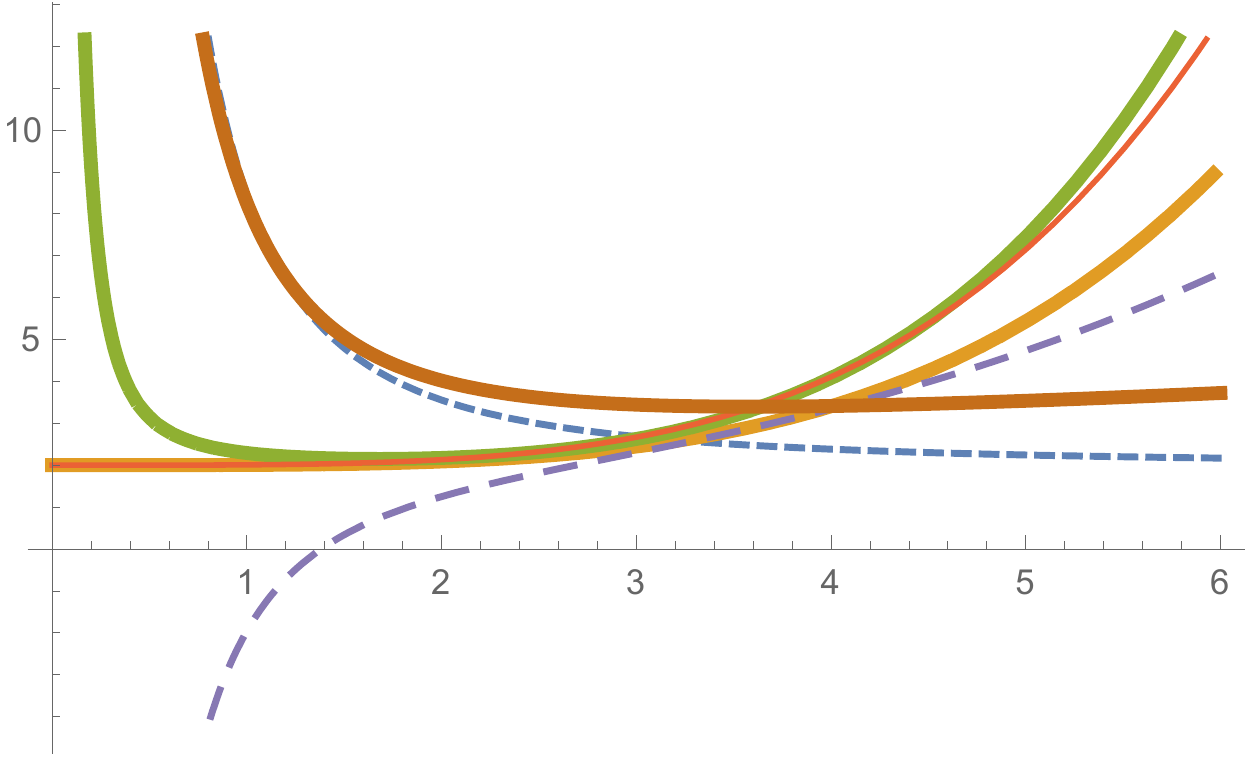}
    \caption{Weak energy condition ($\rho - p$) versus radius ($r$).}
\end{figure}

\newpage
\begin{figure}
    \centering
  \includegraphics[width=6cm]{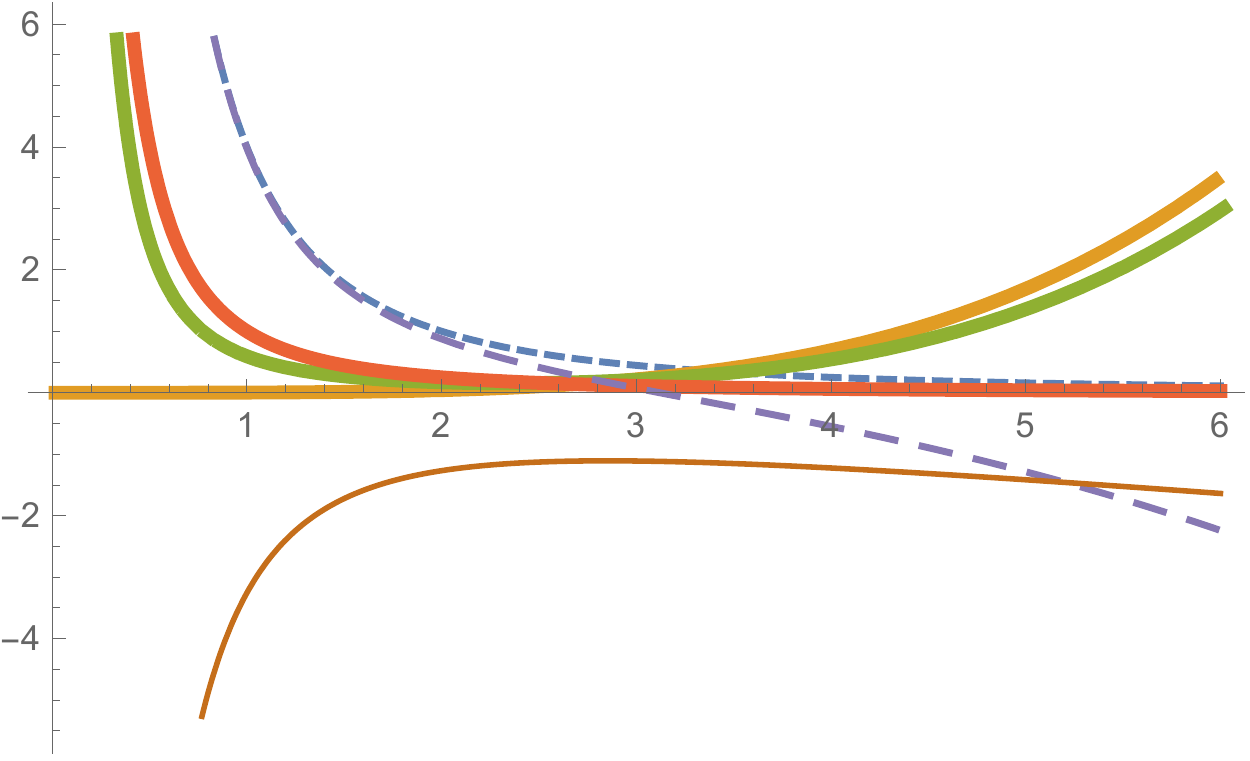}
    \caption{Strong energy condition ($\rho + p$) versus radius ($r$).}
\end{figure}

\newpage
\begin{figure}
    \centering
  \includegraphics[width=6cm]{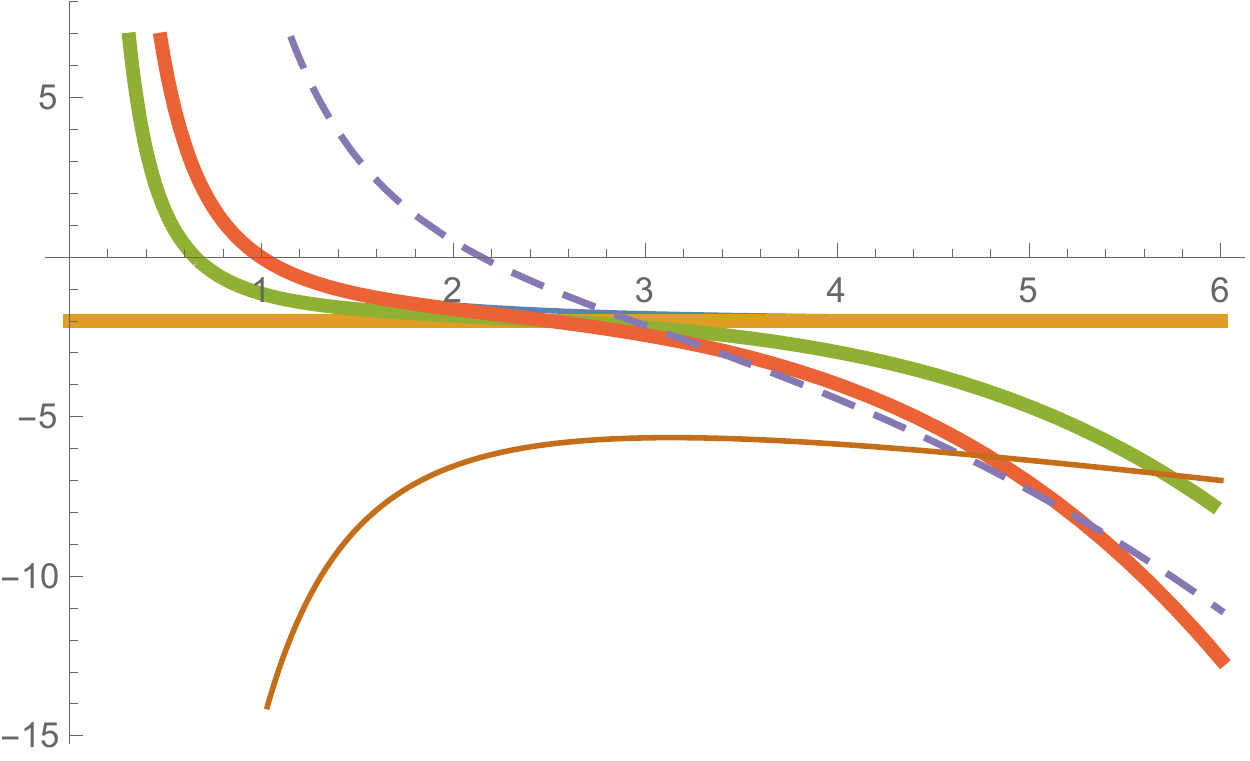}
    \caption{Dominant energy condition ($\rho + 3p$) versus radius ($r$).}
\end{figure}

\newpage
\begin{figure}
    \centering
  \includegraphics[width=6cm]{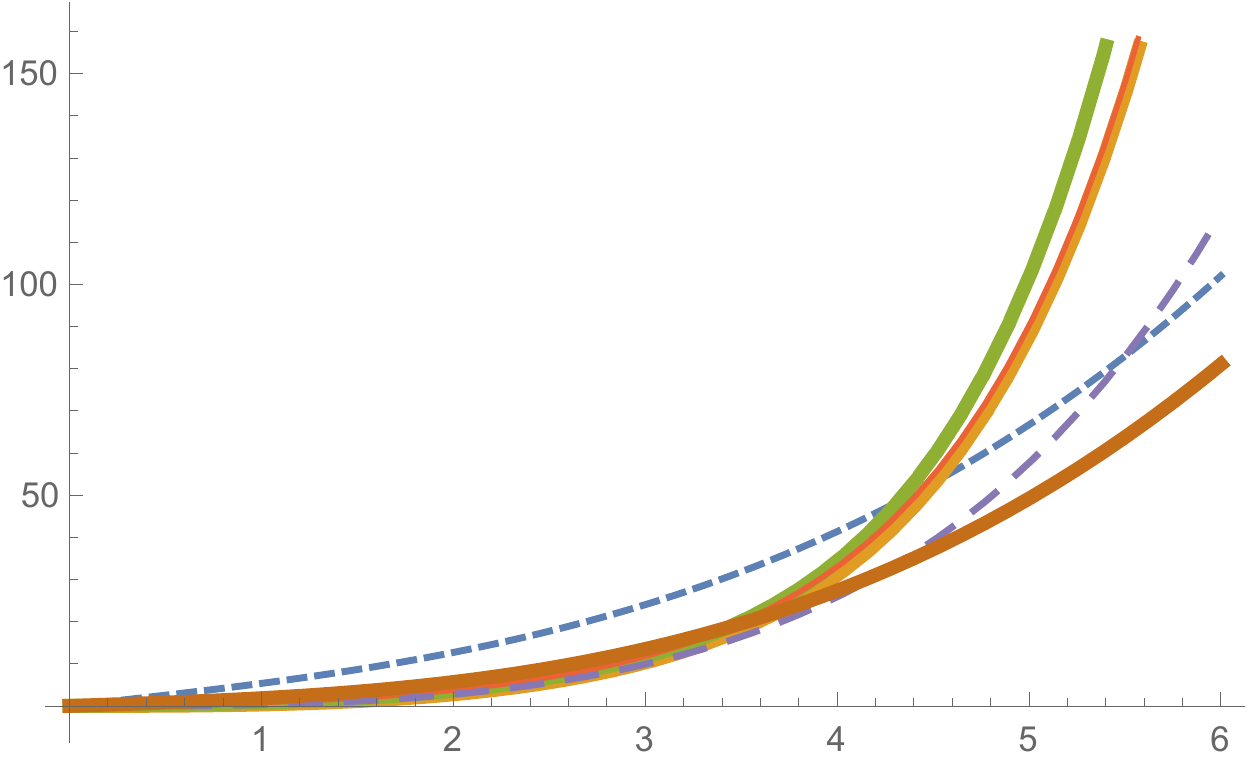}
    \caption{Gravitational mass ($m(r)$) versus radius ($r$).}
\end{figure}

\newpage
\begin{figure}
    \centering
 \includegraphics[width=6cm]{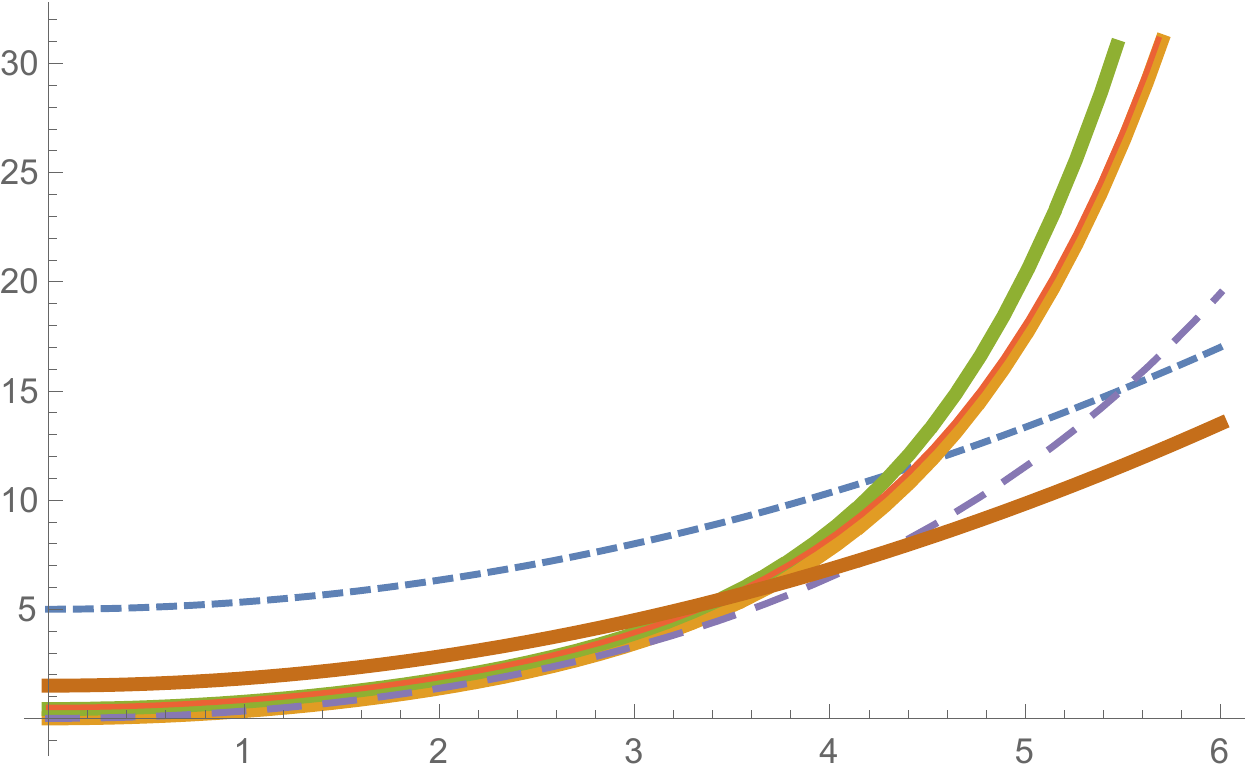}
    \caption{Compactification parameter ($\Lambda$) versus radius ($r$).}
\end{figure}

{\bf Physical Analysis}: The plots represent the physical quantities for various values of $n$. From Fig. 8 it is clear that the energy density is positive for all values of $n$ selected. The pressure (Fig. 9) vanishes in the cases $n = \frac{1}{2}$ at about $r = 0.4$,  $n = 1$ at $r = 0.65$ and $n = 2$ at $r = 1.8$ units. The remaining plots  may now be ignored as they do not meet the basic requirement of a finite radius. Interestingly the $n = \frac{1}{2}$ case (Fig. 10)  meets the causality criterion within the bounded distribution while the case $n = 1$ appears demonstrate the extreme square of sound speed $\frac{dp}{d\rho} = 1$ which is characteristic of stiff fluid matter. The case $ n = 2$ does not support a subluminal sound speed and may now be eliminated from the analysis. The case $n = \frac{1}{2}$ satisfied all the energy requirements (Figs. 11, 12 and 13) within its radius while the case $n=2$ violates the dominant energy condition in the interval $1 < r < 1.8$. Finally the profile of the mass (Fig. 14) conforms to what is expected for all cases of $n$:  that is a smooth increasing function to the boundary layer. The plot of the compactification parameter $\Lambda$ (Fig. 15) suggests that the case $n = \frac{1}{2}$ does satisfy the Buchdahl limit $\frac{m}{r} < \frac{4}{9}$ everywhere within the sphere and specifically at the boundary. Therefore we may conclude that the case $n = \frac{1}{2}$ meets the elementary requirements for physical plausibility. This happens to be the only case studied by Tolman in detail.

\section{Generalized Tolman VI metric}

The choice
$
    e^{-\lambda}=\text{const.}=\frac{1}{2-n^2}
$ suggests  metric potentials $ e^{\lambda}= 2-n^2 \hspace{0.5cm}
{\mbox{and}} \hspace{0.5cm}  e^{\nu}=\left(A r^{1-n}-B
r^{n+1}\right)^2$ for some constants $n$, $A$ and $B$. Within the
trace-free framework we obtain
\begin{eqnarray}
\rho&=&\frac{n^2-1}{\left(n^2-2\right) r^2}-K \\ \nonumber \\
p &=&K-\frac{(n+1)^2-\frac{4 A n}{A-B r^{2 n}}}{\left(n^2-2\right)
r^2}.
\end{eqnarray}
for the dynamical variables. These correspond to Tolman's
calculation and as usual an extra additive constant is in
attendance. The sound speed index has the form
\begin{equation}
\frac{dp}{d\rho}=-\frac{\left(A (n-1)+B (n+1) r^{2
n}\right)^2}{\left(n^2-1\right) \left(A-B r^{2 n}\right)^2}.
\end{equation}
while the energy conditions are governed by the expressions
\begin{eqnarray}
\rho - p&=&\frac{2 n \left(A (n-1)-B (n+1) r^{2 n}\right)}{\left(n^2-2\right) r^2 \left(A-B r^{2 n}\right)}-2 K \\ \nonumber \\
\rho + p&=&\frac{2 \left(A (n-1)+B (n+1) r^{2 n}\right)}{\left(n^2-2\right) r^2 \left(A-B r^{2 n}\right)} \\ \nonumber \\
\rho +3p&=&\frac{2 B (n+1) (n+2) r^{2 n}-2 A (n-2)
(n-1)}{\left(n^2-2\right) r^2 \left(A-B r^{2 n}\right)}+2 K.
\end{eqnarray}
The equation of state may be explicitly determined as
\begin{equation}
p(\rho)=K-\frac{(K+\rho )}{n^2-1} \left((n+1)^2-\frac{4 A n}{A-B
\left(\frac{\sqrt{n^2-1}}{\sqrt{\left(n^2-2\right) (K+\rho
)}}\right)^{2 n}}\right).
\end{equation}
for all values of $n$. Note that the new constant $K$ arising from
the unimodular approach exerts some influence on the equation of
state.  We have found all the necessary quantities to analyse the
model. However, the Tolman VI solution has been criticized for being
irregular. For example, there are singularities in the metric and
dynamical quantities for $n=\pm \sqrt{2}$ and for
$r=\sqrt{\left(\frac{B}{A}\right)^2}$.

 While Tolman utilised the
form $e^{-\lambda}=2-n^2$, the simple prescription $
   e^{\lambda}=\beta
$ for a constant $\beta$ allows us to obtain the remaining potential
as $
    e^{\nu}= \frac{\left(r^{2 \alpha}+c_1\right)^{2}c_2}{r^{2 \left(\alpha-1\right)}}
$ where $\alpha=\sqrt{2-e^{\beta }}$, $\beta$, $c_1$ and $c_2$ are
constants. In this notation the trace--free algorithm results in
\begin{eqnarray}\label{4b2}
    p&=&\frac{e^{-\beta}}{r^2} \left(3+2 \alpha-\frac{4 \alpha c_1}{r^{2 \alpha}+c_1}\right)-\frac{1}{r^2}+K \\ \nonumber \\
    \rho&=&\frac{\sinh \beta-\cosh \beta +1}{r^2}-K \label{4b3}
\end{eqnarray}
for the density and pressure and where $K$ is a constant of
integration. The sound speed has the form
\begin{equation}
    \frac{dp}{d\rho}=\frac{2 \left(e^\beta-1\right) c_1 r^{2 \alpha}-\left(2 \alpha+e^\beta-3\right) c_1^2+\left(2 \alpha-e^\beta+3\right) r^{4 \alpha}}{\left(e^\beta-1\right) \left(r^{2 \alpha}+c_1\right){}^2}.
\end{equation}
while the energy conditions expressions are
\begin{eqnarray}\label{4b5a}
\rho - p&=&\frac{e^{-\beta}}{r^2} \left(\frac{4 \alpha c_1}{r^{2 \alpha}+c_1}-2 \alpha-3\right)+\frac{2}{r^2}+\frac{\sinh \beta-\cosh \beta}{r^2}-2 K\\ \n \\
\rho + p&=&\frac{e^{-\beta}}{r^2} \left(3+2 \alpha-\frac{4 \alpha
c_1}{r^{2 \alpha}+c_1}\right)+\frac{\sinh \beta-\cosh \beta}{r^2}\\
\n \\ \label{4b5b} \rho +3p&=&\frac{2e^{-\beta}}{r^2}
\left(4+3\alpha-\frac{6 \alpha c_1}{r^{2
\alpha}+c_1}\right)-\frac{2}{r^2}+2K. \label{4b5c}
\end{eqnarray}
 Kuchowicz \cite{kuch} exploited a similar approach to
examine the Tolman VI solution and found the general solution as
above but with $K = 0$. On account of the many singularities present
in this model we do not carry out any further study of this case. It is worth noting that $K=0$ results in an inverse square law fall-off of the density and the other constants may also be suitably picked so that the pressure has this same behaviour. In this case isothermal fluid spheres result \cite{saslaw}. 

\section{Extension of the Tolman VII metric}

Commencing with the ansatz $e^{-\lambda}= 1-\frac{r^2}{R^2} +
\frac{4r^4}{A^4} $ the remaining  metric potential evaluates to $
e^\nu= B^2\left[\sin \left( \log \left(\frac{e^{-\frac{\lambda}{2}}
+ 2r^{2}/A^{2}-A^{2}/4R^{2}}{C} \right)^{\frac{1}{2}} \right)
\right]^2 $ as per Tolman. Despite the polynomial assumption
 imposed by  Tolman, the dynamical quantities  became unwieldy.  For this reason, they were omitted in his work. More
importantly the form for $\nu$ is not the most general. To find the
expanded solution, we substitute $\lambda$ from the ansatz into the isotropy
equation (\ref{211a}) and obtain
\begin{equation}
    e^{\nu}=c_2\left(c_1\cos{f}+\sin{f}\right)^2\label{4c4}
\end{equation}
where $c_1$ and $c_2$ are integration constants and \\ $f=\log \sqrt{4
R \left(\sqrt{A^4 \left(R^2-r^2\right)+4 r^4 R^2}+2 r^2
R\right)-A^4}$. The density and pressure are found to be
\begin{eqnarray}
\rho&=&\frac{3}{R^2}-\frac{20r^2}{A^4} \label{4c5}\\ \nonumber \\
p&=&e^{-\lambda} \left(\frac{4 R \left(\cot f -c_1\right)}{\sqrt{A^4
\left(R^2-r^2\right)+4 r^4 R^2} \left(c_1 \cot f
+1\right)}+\frac{1}{r^2}\right)-\frac{1}{r^2} \label{4c5}
\end{eqnarray}
In the case $c_1 = 0$ we regain the original incomplete solution of
Tolman. It is now straightforward but tedious to generate the other
physical quantities however, we omit these lengthy expressions.Note that a barotropic equation of state may be found since $r^2$ may be expressed in terms of $\rho$ and this may be plugged into (\ref{4c5}) to determine $p(\rho)$. 
There is also the prospect of a bounded distribution as it is possible to solve $p(r = R) = 0$.

\section{Tolman VIII metric}

This the last solution Tolman investigated.  Tolman's assumption $
 e^{-\lambda} =\text{const.}r^{-2b}e^{\nu}
$ generates the potentials
$
    e^{-\lambda}=\left(H-\frac{A}{r^n}- \frac{r^q}{F}\right)$ and
    $
    e^{\nu} = B^2r^{2b}e^{-\lambda}
$ where a sequence of constants are defined by  $H=\frac{2}{qn}$ and
$A= (2m)^{n}$; with $q=a-b$, $n=a+2b-1$ and $F$ is a constant.
Within the trace--free framework the  dynamical quantities are
obtained as
\begin{eqnarray}
\rho &=&\left(-b \left((n+2) r^n \left((b-2) F H (q-2)+(2 b+3 q-4)
r^q\right)\right.\right.\nonumber \\&&
 \left.\left.-A F (q-2) (2 b-3 n-4)\right)\right) 
/\left(F (n+2) (q-2)r^{n+2}\right)-K \\ \nonumber \\
p&=&\frac{b}{r^2} \left(\frac{A (n-2 b) }{r^{n}(n+2)}+\frac{(2 b+q)
r^q}{F (q-2)}+b H\right)+K.
\end{eqnarray}
and the familiar integration constant $K$ re-appears. The   sound
speed indicator  is given by
\begin{equation}
\frac{dp}{d\rho}=\frac{A F (n-2 b)+r^n \left(2 b F H-(2 b+q)
r^q\right)}{A F (2 b-3 n-4)+r^n \left((2 b+3 q-4) r^q-2 (b-2) F
H\right)}.
\end{equation}
while the energy conditions are expressed by
\begin{eqnarray}
\rho -p&=&2 \left(-\left((n+2) r^n \left(F (q-2) \left((b-1) b H+K
r^2\right) \right.\right.\right. \nonumber \\ &&
\left.\left.\left.+2 b (b+q-1) r^q\right)-2 A b F
d\right)\right) 
/\left(F (n+2) (q-2)r^{n+2}\right) \\ \nonumber \\
\rho + p&=&\frac{2 b  \left(r^n \left(F H-r^q\right)-A F\right)}{Fr^{n+2}} \\ \nonumber \\
\rho +3p &=&2  \left((n+2) r^n \left(F (q-2) \left(b (b+1) H+K
r^2\right) \right. \right.  \nonumber \\ && \left.\left.+2 b (b+1)
r^q\right)-2 A bF^2 (b+1) (q-2)\right)\nonumber \\&&
 /\left((n+2) (q-2)r^{n+2}\right)
\end{eqnarray}
where $d=(q-2) (b-n-1)$.\\

We observe that in this case the solution given by Tolman is
recovered except for the new constant $K$. A number of special cases
covered earlier may be regained for certain values of $n$ and $q$.
The equation of state may also be determined in some special cases,
however, it cannot be found in general. Accordingly we neglect a
complete study of this case although the additive constant $K$
introduced through the unimodular approach, will have some effect on
the dynamics.

\section{Discussion}

The trace--free Einstein field equations offer an alternative route
to establish the energy density and pressure relevant to a perfect
fluid distribution. Removing the trace of the energy momentum
destroys the energy conservation property however, the coupling of
density and pressure reduces the number of independent field
equations by one. To accommodate this the conservation law may be
added on. We examine the impact on following this presentation of
the field equations by investigating the well known Tolman metrics.
It is found that an additional constant of integration always
appears and this has some bearing on the dynamical behaviour of the
fluid. In the case of the dynamical quantities and energy
conditions, a mere shift is introduced, however, in the active
gravitational mass and equation of state a significant contribution
emerges. We have analysed some of these models with the aid of
graphical plots. In some cases we have extended and generalised the
incomplete solutions provided by Tolman.



\bibliography{basename of .bib file}

\end{document}